\documentclass[twocolumn,prl,preprintnumbers]{revtex4-1}
\usepackage{amsmath,amssymb,graphicx,booktabs,bm,psfrag,color,slashed,euscript,mathtools}
\allowdisplaybreaks
\newcommand{\spac}{{\hspace{0.3mm}}}

\begin{document}

\preprint{IPPP/20-82, MITP/21-007, ZU-TH~01/21}

\title{Consistent treatment of axions in the weak chiral Lagrangian}

\author{Martin Bauer$^a$}
\author{Matthias Neubert$^{b,c,d}$}
\author{Sophie Renner$^e$}
\author{Marvin Schnubel$^b$}
\author{Andrea Thamm$^f$}

\affiliation{${}^a$Institute for Particle Physics Phenomenology, Department of Physics, 
Durham University, Durham, DH1 3LE, UK\\ 
${}^b$PRISMA$^+$\! Cluster of Excellence {\rm \&} MITP, Johannes Gutenberg University, 55099 Mainz, Germany\\
${}^c$Department of Physics {\em\&} LEPP, Cornell University, Ithaca, NY 14853, U.S.A.\\
${}^d$Department of Physics, Universit\"at Z\"urich, Winterthurerstrasse 190, CH-8057 Z\"urich, Switzerland\\
${}^e$SISSA International School for Advanced Studies, Via Bonomea 265, 34136, Trieste, Italy\\
${}^f$School of Physics, The University of Melbourne, Victoria 3010, Australia}

\begin{abstract}
We present a consistent implementation of weak decays involving an axion or axion-like particle in the context of an effective chiral Lagrangian. We argue that previous treatments of such processes have used an incorrect representation of the flavor-changing quark currents in the chiral theory. As an application, we derive model-independent results for the decays $K^-\to\pi^- a$ and $\pi^-\to e^-\bar\nu_e\spac a$ at leading order in the chiral expansion and for arbitrary axion couplings and mass. In particular, we find that the $K^-\to\pi^- a$ branching ratio is almost 40 times larger than previously estimated.  
\end{abstract}

\maketitle

Axions and axion-like particles (collectively referred to as ALPs in this work) are new types of elementary particles, which arise in a large class of extensions of the Standard Model (SM) and are well motivated theoretically. They can provide an elegant solution to the strong CP problem based on the Peccei--Quinn mechanism \cite{Peccei:1977hh,Weinberg:1977ma,Wilczek:1977pj,Bardeen:1977bd,Kim:1979if,Shifman:1979if,Dine:1981rt,Zhitnitsky:1980tq}. More generally, ALPs can arise as pseudo Nambu--Goldstone bosons in models with explicit global symmetry breaking. Low-energy weak-interaction processes imply some of the most stringent bounds on the couplings of ALPs to gluons and other SM particles \cite{Bardeen:1978nq,Antoniadis:1981zw,Krauss:1986bq,Bardeen:1986yb}.

In a seminal paper \cite{Georgi:1986df}, Georgi, Kaplan and Randall have derived the effective chiral Lagrangian accounting for the interactions of a light ALP (with mass below the scale of chiral symmetry breaking, $\mu_\chi=4\pi f_\pi$) with the light pseudoscalar mesons, opening the door to a model-independent description which does not rely on the details of Peccei--Quinn symmetry breaking. In this Letter, we reanalyze this problem and point out a small but important omission in the representation of the weak-interaction quark currents in the effective theory, which has far-reaching consequences. Despite the 35-year history of the subject, we find that even recent papers on weak decays such as $K^-\to\pi^- a$ and $\pi^-\to e^-\bar\nu_e\spac a$ omit the contributions of relevant Feynman diagrams and thus employ incomplete  expressions for the decay amplitudes (see e.g.\ \cite{Bjorkeroth:2018dzu,Ertas:2020xcc,Gori:2020xvq}). In many phenomenological studies, the amplitudes are derived by starting from an amplitude for a decay process involving a $\pi^0$ or $\eta$ meson and accounting for the (kinetic) mixing of the ALP with these neutral mesons by means of mixing angles $\theta_{\pi a}$ and $\theta_{\eta a}$. Below we recall the well-known fact that in the approach of \cite{Georgi:1986df} the mixing angles are unphysical, because they depend on the parameters of the chiral rotation used to eliminate the ALP--gluon coupling in the effective Lagrangian. It is customary to adopt a ``default choice'' for these parameters, which eliminates the mass mixing in the effective Lagrangian. However, there always exist other contributions to the decay amplitude, in which the ALP participates in the relevant interaction vertices. Neglecting these ``direct'' contributions leads to incorrect predictions. In fact, they are essential to ensure that the auxiliary parameters of the chiral rotation cancel out in predictions for physical quantities. (Only a very special class of models, in which the ALP couples to SM fields only through phases in the quark mass matrices, with no derivative interactions and no couplings to gluons at the low scale $\mu_\chi$, is an exception to this rule, see e.g.\ \cite{Krauss:1986bq,Alves:2017avw}.) 

The starting point of our study is the effective ALP Lagrangian at a scale of order $\mu_\chi\approx 1.6$\,GeV, which we write in the form \cite{Georgi:1986df}
\begin{equation}\label{Leff}
\begin{aligned}
   {\cal L}_{\rm eff}
   &= {\cal L}_{\rm QCD} + \frac12 \left( \partial_\mu a\right)\!\left( \partial^\mu a\right) 
    - \frac{m_{a,0}^2}{2}\,a^2 \\[0.5mm]
   &\quad + c_{GG}\,\frac{\alpha_s}{4\pi}\,\frac{a}{f}\,G_{\mu\nu}^a\,\tilde G^{\mu\nu,a}
    + c_{\gamma\gamma}\,\frac{\alpha}{4\pi}\,\frac{a}{f}\,F_{\mu\nu}\,\tilde F^{\mu\nu} \\
   &\quad + \frac{\partial^\mu a}{f}\,\Big( \bar q_L\spac\bm{k}_Q\spac\gamma_\mu\spac q_L
    + \bar q_R\,\bm{k}_q\spac\gamma_\mu\spac q_R + \dots \Big) \,.
\end{aligned}
\end{equation}
Here $q$ is a 3-component vector in generation space containing the three light quark flavors $u,d,s$. The ALP decay constant $f$ is related to the scale of global (Peccei--Quinn) symmetry breaking by $\Lambda=4\pi f$ and is assumed to lie above the scale of electroweak symmetry breaking. It governs the overall magnitude of the ALP interactions with SM particles, the leading of which are mediated by dimension-5 operators. (In the literature on QCD axions, one often defines the axion decay constant $f_a$ in terms of the strength of the axion--gluon coupling, such that $1/f_a=-2c_{GG}/f$.) The parameters $c_{GG}$ and $c_{\gamma\gamma}$ determine the strengths of the ALP interactions with gluons and photons, while the hermitian matrices $\bm{k}_Q$ and $\bm{k}_q$ contain the ALP couplings to left-handed and right-handed quarks. The off-diagonal entries of these matrices account for the possibility of flavor-changing $s\to d$ transitions. The dots represent analogous couplings to leptons. The ALP couplings are scale-dependent quantities. Their evolution from the new-physics scale $\Lambda$ down to the scale $\mu_\chi$ has recently been studied in detail \cite{Chala:2020wvs,Bauer:2020jbp}. The mass parameter $m_{a,0}^2$ provides an explicit soft breaking of the shift symmetry $a\to a+c$, which is a (classical) symmetry of the effective Lagrangian (\ref{Leff}). In QCD axion models $m_{a,0}^2$ vanishes and the axion mass is generated by non-perturbative QCD dynamics \cite{Shifman:1979if,DiVecchia:1980yfw}. In more general ALP models a non-zero bare mass can be generated by means of non-abelian extensions of the SM.  

To study the low-energy interactions of a light ALP with the pseudoscalar mesons $(\pi,K,\eta)$, the Lagrangian (\ref{Leff}) is matched onto a chiral effective Lagrangian, in which $\bm{\Sigma}(x)=\exp\big[\frac{i\sqrt2}{f_\pi}\,\lambda^a\spac\pi^a(x)\big]$ contains the pseudoscalar meson fields ($\lambda_a$ are the Gell-Mann matrices). In order to find the bosonized form of the ALP--gluon interaction, one eliminates the $a\spac G\tilde G$ term in favor of ALP couplings to quark bilinears, whose chiral representation is well known. This is accomplished with a chiral rotation \cite{Georgi:1986df,Bardeen:1986yb,Srednicki:1985xd} 
\begin{equation}\label{chiralrot}
   q(x)\to \exp\left[ -i\left( \bm{\delta}_q + \bm{\kappa}_q\spac\gamma_5 \right) c_{GG}\,\frac{a(x)}{f} 
    \right] q(x) \,, 
\end{equation}
where $\bm{\delta}_q$ and $\bm{\kappa}_q$ are hermitian matrices, which we choose to be diagonal in the quark mass basis. Under this field redefinition the measure of the path integral is not invariant \cite{Fujikawa:1979ay}, and this generates extra contributions to the ALP couplings to gluons and photons. Imposing the condition 
\begin{equation}\label{trace}
   \mbox{Tr}\,\bm{\kappa}_q = \kappa_u + \kappa_d + \kappa_s = 1
\end{equation}
ensures that the ALP--gluon interaction is eliminated from the Lagrangian at the expense of modifying the ALP--photon and ALP--fermion couplings as well as the quark mass matrix. Denoting the modified couplings with a hat, one finds (with $N_c=3$ the number of colors)
\begin{equation}
\begin{aligned}
   \hat{c}_{\gamma\gamma} 
   &= c_{\gamma\gamma} - 2N_c\,c_{GG}\,\text{Tr}\,\bm{Q}^2\spac\bm{\kappa}_q \,, \\
   \hat{\bm{k}}_Q(a) 
   &= e^{i\bm{\phi}_q^- a/f} \left( \bm{k}_Q + \bm{\phi}_q^- \right) e^{-i\bm{\phi}_q^- a/f} \spac , \\
   \hat{\bm{k}}_q(a) 
   &= e^{i\bm{\phi}_q^+ a/f} \left( \bm{k}_q + \bm{\phi}_q^+ \right) e^{-i\bm{\phi}_q^+ a/f} ,
\end{aligned}
\end{equation}
where $\bm{\phi}_q^\pm=c_{GG}\spac(\bm{\delta}_q\pm\bm{\kappa}_q)$, and $\bm{Q}=\text{diag}(Q_u,Q_d,Q_s)$ contains the electric charges of the quarks in units of $e$. The phase factors in the last two relations cancel for all diagonal elements of the matrices $\hat{\bm{k}}_Q$ and $\hat{\bm{k}}_q$. As long as the condition (\ref{trace}) is satisfied, any choice of the matrices $\bm{\delta}_q$ and $\bm{\kappa}_q$ describes the same physics. The derivative couplings of the ALP to the left- and right-handed quark currents are implemented by including the ALP field in the definition of the covariant derivative \cite{Gasser:1984gg}, such that
\begin{equation}\label{eq:5}
   i\bm{D}_\mu\bm{\Sigma} 
   = i\partial_\mu\bm{\Sigma} + e\spac A_\mu\spac[\bm{Q},\bm{\Sigma}]
    + \frac{\partial_\mu a}{f} \left( \hat{\bm{k}}_Q\spac\bm{\Sigma} 
    - \bm{\Sigma}\spac\spac\hat{\bm{k}}_q \right) , 
\end{equation}
where $A_\mu$ is the photon field. This definition implies
\begin{equation}
   (\bm{D}_\mu\bm{\Sigma})\,\bm{\Sigma}^\dagger
    + \bm{\Sigma}\,(\bm{D}_\mu\bm{\Sigma})^\dagger
   = \partial_\mu\!\left( \bm{\Sigma}\,\bm{\Sigma}^\dagger \right) 
   = 0 \,.
\end{equation}
The leading-order chiral Lagrangian can then be expressed in the form
\begin{equation}\label{chiPT}
\begin{aligned}
   {\cal L}_{\rm eff}^\chi
   &= \frac{f_\pi^2}{8}\spac\mbox{Tr}\big[ \bm{D}^\mu\bm{\Sigma}\,(\bm{D}_\mu\bm{\Sigma})^\dagger \big] 
    + \frac{f_\pi^2}{4}\spac B_0\spac\mbox{Tr}\big[ \hat{\bm{m}}_q(a)\spac\bm{\Sigma}^\dagger 
    \!+\hspace{-0.2mm} \text{h.c.} \big] \\
   &\quad + \frac12\,\partial^\mu a\,\partial_\mu a - \frac{m_{a,0}^2}{2}\,a^2
    + \hat c_{\gamma\gamma}\,\frac{\alpha}{4\pi}\,\frac{a}{f}\,F_{\mu\nu}\,\tilde F^{\mu\nu} \spac ,
\end{aligned}
\end{equation} 
where the parameter $B_0\approx m_\pi^2/(m_u+m_d)$ is proportional to the chiral condensate. Throughout this Letter we work consistently at lowest order in the chiral expansion and neglect the effects of $\pi^0$--\spac$\eta$\spac--\spac$\eta'$ mixing. With our choice of diagonal matrices $\bm{\delta}_q$ and $\bm{\kappa}_q$, the modified quark mass matrix takes the form
\begin{equation}\label{mhatq}
   \hat{\bm{m}}_q(a) 
   = \exp\left( - 2i\bm{\kappa}_q\spac c_{GG}\,\frac{a}{f} \right) \bm{m}_q \,,
\end{equation}
where $\bm{m}_q=\text{diag}(m_u,m_d,m_s)$.

The effective chiral Lagrangian (\ref{chiPT}) has been the basis for numerous studies of low-energy phenomena involving axions or light ALPs. Expanding the Lagrangian to quadratic order in fields, one finds that the ALP acquires the mass term
\begin{equation}
   m_a^2 = c_{GG}^2\,\frac{f_\pi^2\,m_\pi^2}{f^2}\,\frac{2m_u\spac m_d}{(m_u+m_d)^2} 
    + m_{a,0}^2 \left[ 1 + {\cal O}\bigg(\frac{f_\pi^2}{f^2}\bigg) \right] ,
\end{equation} 
up to higher-order corrections in the chiral expansion \cite{Shifman:1979if,DiVecchia:1980yfw}. Higher-order terms generate a periodic potential for the ALP field $a$, which breaks the continuous shift symmetry of the classical Lagrangian to the discrete shift symmetry $a\to a+n\pi f/c_{GG}$. One also finds that there are mass-mixing and kinetic-mixing contributions involving the ALP and the neutral mesons $\pi^0$ and $\eta$, whose explicit form depends on the parameters $\kappa_q$. For instance, at first order in $1/f$ one obtains $\pi^0=\pi_{\rm phys}^0+\theta_{\pi a}\spac a_{\rm phys}$ with the mixing angle
\begin{equation}\label{eq:10}
   \theta_{\pi a} 
   = \frac{f_\pi}{2\sqrt2\spac f} \left[
    \frac{m_a^2\spac(\hat c_{uu}-\hat c_{dd})}{m_\pi^2-m_a^2} 
    - \frac{m_\pi^2\spac\Delta_\kappa}{m_\pi^2-m_a^2} \right] ,
\end{equation} 
where $\hat c_{qq}=c_{qq}+2\kappa_q\spac c_{GG}$ with 
\begin{equation}
   c_{qq} = (k_q-k_Q) \,, \quad
   \Delta_\kappa = 4 c_{GG}\,\frac{m_u\spac\kappa_u-m_d\spac\kappa_d}{m_d+m_u} \,.
\end{equation}
Via the quantities $\hat c_{qq}$ and $\Delta_\kappa$ the mixing angle depends on the auxiliary parameters $\kappa_q$ in (\ref{chiralrot}). The special choice $\bm{\kappa}_q=\bm{m}_q^{-1}/\text{Tr}\spac(\bm{m}_q^{-1})$ eliminates the mass-mixing contribution $\Delta_\kappa$, leaving a contribution from kinetic mixing that is proportional to $m_a^2$ and hence is negligible for a QCD axion with $m_a^2\sim f_\pi^2/f^2$. This ``default choice'' defines a scheme, which is frequently adopted in the literature. It is important to realize, however, that $\theta_{\pi a}$ is not a physical quantity. For instance, one can find values of $\kappa_u$, $\kappa_d$ and $\kappa_s$ such that $\theta_{\pi a}=0$ and $\theta_{\eta a}=0$ \cite{Bauer:2020jbp}. In our discussion below we treat the quantities $\delta_q$ and $\kappa_q$ in the field redefinition (\ref{chiralrot}) as free parameters, subject only to condition (\ref{trace}). We study in detail how the dependence on these auxiliary variables cancels in predictions for physical observables. For flavor-conserving processes such as $a\to\gamma\gamma$ and $a\to\pi\pi\pi$, an analogous study was performed in \cite{Bauer:2020jbp}.

In (\ref{chiPT}) the ALP enters in the quark mass matrix $\hat{\bm{m}}_q(a)$ and through the covariant derivative defined in (\ref{eq:5}). For the very special situation in which 
\begin{equation}\label{specialcond}
   \text{Tr}\spac\big[\bm{k}_Q(\mu_\chi) - \bm{k}_q(\mu_\chi)\big] = 2c_{GG} \,, 
\end{equation}
it is possible to choose the matrices $\bm{\kappa}_q$ and $\bm{\delta}_q$ in such a way that $\hat{\bm{k}}_q$ and $\hat{\bm{k}}_Q$ both vanish. In this case, the ALP only enters the Lagrangian through the quark mass matrix (\ref{mhatq}), see e.g.\ \cite{Alves:2017avw}. However, condition (\ref{specialcond}) is not invariant under renormalization-group evolution, and it would need a fine tuning to realize this condition at the low scale $\mu_\chi$. 

The effective chiral Lagrangian (\ref{chiPT}) can also be used to study flavor-changing processes such as $K^-\to\pi^- a$ and $\pi^-\to e^-\bar\nu_e\spac a$, which in the SM are mediated by the weak interactions and at low energies are described by 4-fermion operators built out of products of left-handed currents. Under a left-handed, flavor off-diagonal rotation $q_L\to\bm{U}_L\spac q_L$ of the quark fields, the meson fields transform non-linearly as $\bm{\Sigma}\to\bm{U}_L\spac\bm{\Sigma}$. The effective Lagrangian is invariant under this transformation if we treat the quark mass matrix and the left-handed ALP couplings as spurions transforming as $\hat{\bm{m}}_q(a)\to\bm{U}_L\spac\hat{\bm{m}}_q(a)$ and $\hat{\bm{k}}_Q\to\bm{U}_L\spac\hat{\bm{k}}_Q\spac\bm{U}_L^\dagger$. Applying the Noether procedure to the Lagrangians in the quark and meson pictures, and accounting for an additional phase factor arising from the chiral rotation of the fields, we find that the left-handed quark currents $\bar q_L^i\gamma_\mu q_L^j$ must be represented in the chiral theory by
\begin{align}\label{master}
   & L_\mu^{ji} 
    = - \frac{i f_\pi^2}{4}\,e^{i(\phi_{q_i}^- -\phi_{q_j}^-)\spac a/f} 
    \big[ \bm{\Sigma}\,(\bm{D}_\mu\bm{\Sigma})^\dagger \big]^{ji} \notag\\
   &\ni - \frac{i f_\pi^2}{4} \left[ 1 
    + i(\delta_{q_i}-\delta_{q_j}-\kappa_{q_i}+\kappa_{q_j})\,c_{GG}\spac\frac{a}{f} 
    \right] \big[ \bm{\Sigma}\,\partial_\mu\bm{\Sigma}^\dagger \big]^{ji} \notag\\
   &\quad + \frac{f_\pi^2}{4}\,\frac{\partial^\mu a}{f}\,\big[ \hat{\bm{k}}_Q
    - \bm{\Sigma}\,\hat{\bm{k}}_q\spac\bm{\Sigma}^\dagger \big]^{ji} \,.
\end{align} 
This generates both non-derivative and derivative couplings of the ALP to the weak-interaction vertices. With the special choice $\delta_q=\kappa_q$ one can eliminate the non-derivative couplings; however, the derivative couplings remain. Astoundingly, it appears that the contribution involving the derivative of the ALP field has been omitted in the literature. It has neither been taken into account in the original paper \cite{Georgi:1986df} nor in later work based on it.

\begin{figure}[t]
\begin{center}
\vspace{1mm}
\includegraphics[width=0.48\textwidth]{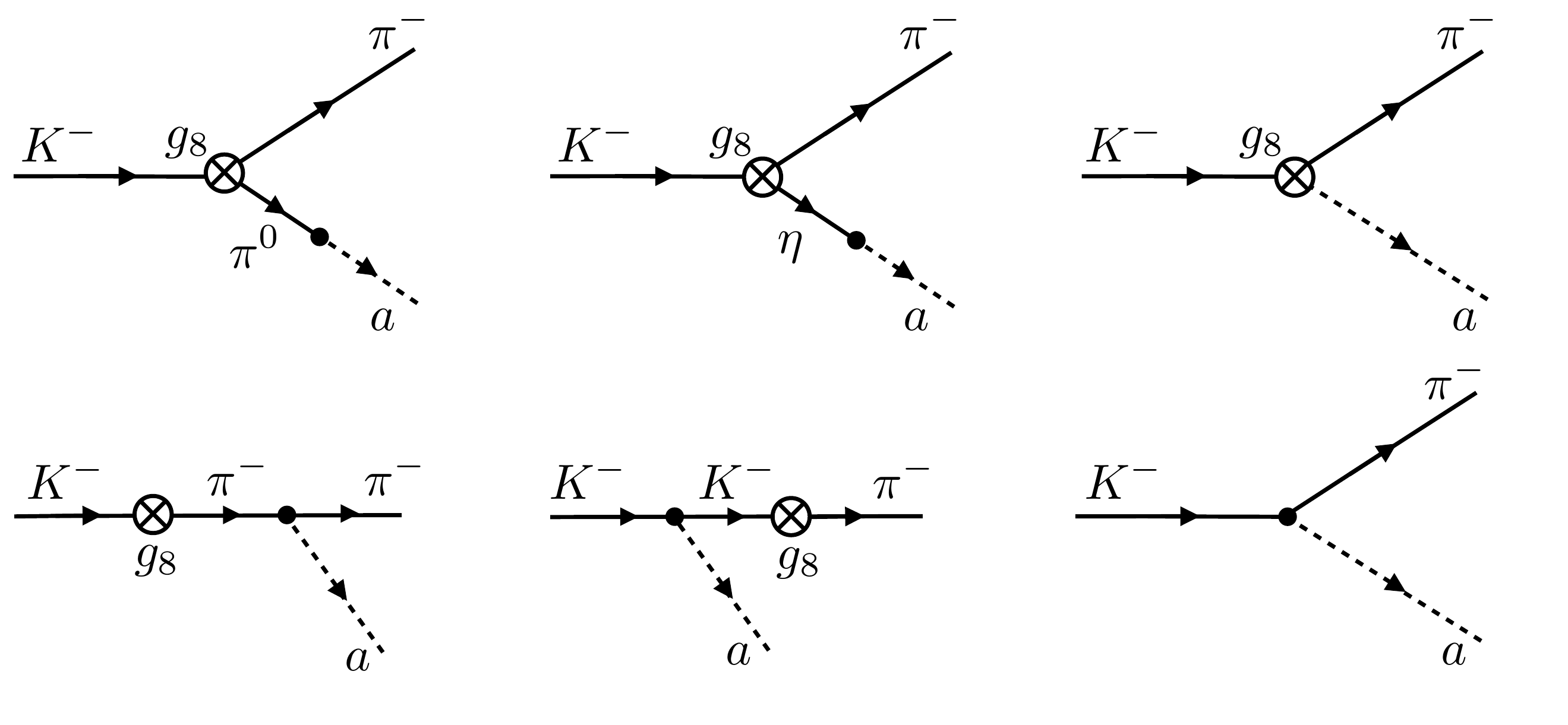} 
\caption{\label{fig:graphs} 
Feynman graphs contributing to the $K^-\to\pi^- a$ decay amplitude at leading order in the chiral expansion. Weak-interaction vertices are indicated by a crossed circle, while dots refer to vertices from the Lagrangian (\ref{chiPT}).}
\vspace{-6mm}
\end{center}
\end{figure}

The chiral representation of the effective weak Lagrangian mediating the decays $K^-\to\pi^-\pi^0$, $K_S\to\pi^+\pi^-$ and $K_S\to\pi^0\pi^0$ at leading order in the chiral expansion involves an operator transforming as an $SU(3)$ octet and two transforming as 27-plets \cite{Bernard:1985wf,Crewther:1985zt,Kambor:1989tz}. (A second octet operator can be transformed into the first one using the equations of motion.) The octet operator receives a huge dynamical enhancement known as the $\Delta I=\frac12$ selection rule \cite{Neubert:1991zd}. The corresponding Lagrangian reads
\begin{equation}\label{Lweak}
   {\cal L}_{\rm weak}
   = - \frac{4G_F}{\sqrt2}\,V_{ud}^*\spac V_{us}\,g_8 \left[ L_\mu\spac L^\mu \right]^{32} \spac ,
\end{equation}
where $|g_8|\approx 5.0$ \cite{Cirigliano:2011ny}, and the index pair ``32'' signals a $s_L\to d_L$ transition. We have calculated the $K^-\to\pi^- a$ decay amplitude from the Lagrangians (\ref{chiPT}) and (\ref{Lweak}), evaluating the Feynman graphs shown in Figure~\ref{fig:graphs}. The first two diagrams account for the ALP--meson mixing contributions mentioned above, while the third graph contains the ALP interactions at the weak vertex derived from (\ref{master}). The following two graphs describe ALP emission of an initial- or final-state meson. They give nonzero contributions if the ALP has non-universal vector-current interactions with different quark flavors. The last diagram contains possible flavor-changing ALP--fermion couplings, as parameterized by the off-diagonal elements of the matrices $\bm{k}_Q$ and $\bm{k}_q$ in (\ref{Leff}). To simplify the analysis we set $m_u=m_d\equiv\bar m$ in order to eliminate the $\pi^0$--\spac$\eta$ mass mixing. (More general expressions, including also the contribution from the 27-plet operators, will be presented elsewhere.) The meson masses are then given by $m_\pi^2=2B_0\spac\bar m$, $m_K^2=B_0\spac(m_s+\bar m)$, and $3m_\eta^2=4m_K^2-m_\pi^2$. Before considering the resulting decay amplitude, it is instructive to see how the scheme-dependent contributions involving the $\delta_q$ and $\kappa_q$ parameters cancel between the various diagrams. In units of $N_8=-\frac{G_F}{\sqrt2}\,V_{ud}^*\spac V_{us}\,g_8\spac f_\pi^2$, with $|N_8|\approx 1.53\cdot 10^{-7}$, we find for these contributions
\begin{align}
   D_1 &\ni \frac{N_8}{2f}\,c_{GG}\spac(\kappa_u-\kappa_d)\spac(m_\pi^2-m_a^2) \,, \notag\\
   D_2 &\ni - \frac{N_8}{6f}\,c_{GG}\,(2m_K^2+m_\pi^2-3m_a^2)\,(\kappa_u+\kappa_d-2\kappa_s) \,, \notag\\
   D_3 &\ni \frac{N_8}{2f}\,c_{GG}\,\Big[\! - (\delta_d-\delta_s-\kappa_d+\kappa_s)\spac
    (m_K^2+m_\pi^2-m_a^2) \notag\\[-0.8mm]
   &\hspace{1.77cm} + (\delta_u-\delta_d+\kappa_u+\kappa_s)\spac(m_K^2-m_\pi^2+m_a^2) \notag\\
   &\hspace{1.77cm} + (\delta_u-\delta_s+\kappa_u+\kappa_d)\spac(m_K^2-m_\pi^2-m_a^2) \Big] \spac, \notag\\
   D_4 &\ni - \frac{N_8}{f}\,c_{GG}\,m_K^2\,(\delta_u-\delta_d) \,, \notag\\
   D_5 &\ni \frac{N_8}{f}\,c_{GG}\,m_\pi^2\,(\delta_u-\delta_s) \,, 
\end{align}
while the last diagram is scheme independent. Via the mixing angles $\theta_{\pi a}$ and $\theta_{\eta a}$ the results for $D_1$ and $D_2$ depend on the $\kappa_q$ parameters, see (\ref{eq:10}). The expressions for $D_4$ and $D_5$, on the other hand, depend only on the $\delta_q$ parameters. Only the third diagram, in which the ALP is emitted from the weak-interaction vertex, depends on both sets of parameters. In the sum of all contributions the dependence on the auxiliary parameters cancels (apart from an unambiguous contribution proportional to $\kappa_u+\kappa_d+\kappa_s=1$). But this cancellation only works if the derivative ALP interactions in (\ref{master}) are included. 

Adding up all contributions, we obtain for the decay amplitude (for $m_u=m_d$)
\begin{align}\label{resu}
   & i{\cal A}_{K^-\to\pi^- a} = \frac{N_8}{4f}\,\bigg[
    16\spac c_{GG}\spac\frac{(m_K^2-m_\pi^2)(m_K^2-m_a^2)}{4m_K^2-m_\pi^2-3m_a^2} \notag\\
   &\quad + 6\spac(c_{uu}+c_{dd}-2c_{ss})\,m_a^2\,\frac{m_K^2-m_a^2}{4m_K^2-m_\pi^2-3m_a^2} \notag\\
   &\quad + (2c_{uu}+c_{dd}+c_{ss})\,(m_K^2-m_\pi^2-m_a^2) + 4\spac c_{ss}\spac m_a^2 \notag\\
   &\quad + (k_d+k_D-k_s-k_S)\,(m_K^2+m_\pi^2-m_a^2) \bigg] \notag\\[-1mm]
   &\quad - \frac{m_K^2-m_\pi^2}{2f} \left[\spac k_q+k_Q\right]^{23} .
\end{align}
Note that the transition $K^-\to\pi^- a$ proceeds via the dynamically enhanced octet operator, whereas the corresponding decay $K^-\to\pi^-\pi^0$ receives contributions from the 27-plet operator with isospin change $\Delta I=\frac32$ only. This effect is well known and is referred to as ``octet enhancement'' \cite{Bardeen:1978nq,Antoniadis:1981zw}. Attempts to estimate the $K^-\to\pi^- a$ decay rate as $\theta_{\pi a}^2$ times the $K^-\to\pi^-\pi^0$ rate miss this important effect. Another interesting feature of the result (\ref{resu}) is its dependence on the flavor-conserving ALP vector couplings $(k_d+k_D)$ and $(k_s+k_S)$ to down and strange quarks. In the presence of the weak interactions the currents $\bar d\spac\gamma_\mu d$ and $\bar s\spac\gamma_\mu\spac s$ are not individually conserved (unlike in QCD), and hence these couplings can have observable effects.

In order to compare our result (\ref{resu}) with some previous calculations, we work to leading order in the ratio $\bar m/m_s$, consider the limit where $m_a^2\ll m_K^2$ and assume the case of a minimal flavor-violating ALP, for which $c_{ss}=c_{dd}$ and $k_d+k_D=k_s+k_S$ \cite{Bauer:2020jbp}. We then obtain the simple result (still with $m_u=m_d$, neglecting the small 27-plet contributions, and setting $1/f_a=-2c_{GG}/f$)
\begin{equation}\label{eq:17}
   {\cal A}_{K^-\to\pi^- a} 
   \approx \frac{i m_K^2}{2f_a}\!\left[ N_8\! \left( 1 + \frac{c_{uu}+c_{dd}}{2 c_{GG}} \right) 
    - \frac{\left[\spac k_q+k_Q\right]^{23}}{2c_{GG}} \right]\! .
\end{equation}
Barring cancellations, the contribution proportional to $N_8$ dominates as long as $|[\spac k_q+k_Q]^{23}/c_{GG}|\ll 3\cdot 10^{-7}$, which we assume from now on. Eliminating the parameter $N_8$ via the $K_S\to\pi^+\pi^-$ decay amplitude, we obtain
\begin{equation}
   \frac{\text{Br}(K^-\to\pi^- a)}{\text{Br}(K_S\to\pi^+\pi^-)} 
   \approx \frac{\tau_{K^-}}{\tau_{K_S}}\,
    \frac{f_\pi^2}{8f_a^2} \left[ 1 + \frac{c_{uu}+c_{dd}}{2 c_{GG}} \right]^2 .
\end{equation}
For a long-lived ALP with mass $m_a\ll m_\pi$, the upper limit $\text{Br}(K^-\to\pi^- X)<2.0\cdot 10^{-10}$ (90\% CL) reported by NA62 \cite{CortinaGil:2020fcx} from a search for a feebly interacting new particle $X$ implies  
\begin{equation}\label{fabound}
   \frac{1}{f_a} \left| 1 + \frac{c_{uu}+c_{dd}}{2 c_{GG}} \right|
   < \frac{1}{31.9\,\text{TeV}} \,.
\end{equation}
Estimating the weak-interaction contribution to the decay amplitude from kinetic ALP--meson mixing (see e.g.\ \cite{Bjorkeroth:2018dzu,Ertas:2020xcc,Gori:2020xvq}) corresponds to retaining only the first two diagrams in Figure~\ref{fig:graphs}, evaluated with the default choice of $\kappa_q$ parameters. Under the approximations described above this leads to
\begin{equation}
   {\cal A}_{K^-\to\pi^- a} 
   \approx \frac{i N_8\spac m_a^2}{8f_a} \left( 1 - \frac{c_{uu}-c_{dd}}{2 c_{GG}} \right) ,
\end{equation}
which underestimates the amplitude by a factor $m_a^2/(4m_K^2)$ and predicts the wrong sign for the contribution proportional to $c_{uu}$. If mass mixing with the $\eta'$ is included, one finds an additional small contribution proportional to $\sin\theta_{\eta\eta'}\,m_\pi^2/m_K^2$ \cite{Ertas:2020xcc,Gori:2020xvq} relative to the leading term in our result. The authors of \cite{Georgi:1986df} performed a more careful evaluation of the $K^-\to\pi^- a$ decay rate for the case of a QCD axion ($m_a^2\approx 0$) without couplings to matter ($c_{qq}=0$). In this case diagrams $D_1$ and $D_2$ vanish when one adopts the default choice of $\kappa_q$ parameters, and the graphs $D_4$ and $D_5$ vanish if one chooses $\delta_q=0$. In the evaluation of the third diagram the authors omitted the derivative couplings of the axion shown by the last term in (\ref{master}). They obtained (this formula was not explicitly shown in the paper, but we have derived it from their arguments and the presented numerical result) 
\begin{equation}
   {\cal A}_{K^-\to\pi^- a} 
   \approx \frac{i N_8\spac m_K^2}{4f_a}\,\frac{m_u}{m_u+m_d} \,.
\end{equation}
This contribution to the amplitude is smaller than the corresponding term in (\ref{eq:17}) by a factor $\frac{m_u}{2(m_u+m_d)}\approx 0.16$, corresponding to an underestimation of the branching ratio by about a factor 37. (In \cite{Georgi:1986df} the authors state that they have derived the same result in a different scheme with $\delta_q=\kappa_q$, in which the ALP is removed from the weak-interaction vertex. With their omission, we cannot reproduce that the two treatments lead to the same expression.)

We have also applied our matching prescription (\ref{master}) to derive the $\pi^-\to e^-\bar\nu_e\spac a$ decay amplitude, finding again a result that is independent of the choice of the $\delta_q$ and $\kappa_q$ parameters. It reads
\begin{equation}\label{eq:21}
\begin{aligned}
   & {\cal A}_{\pi^-\to e^-\bar\nu_e\spac a} 
    = - \frac{G_F}{\sqrt2}\,V_{ud}\,\frac{f_\pi}{2f}\,
    \bar u_e\spac(\rlap{\hspace{0.2mm}/}{p_\pi}+\rlap{\hspace{0.2mm}/}{p_a})
    (1-\gamma_5)\,v_{\bar\nu_e} \\
   &\quad \times \left[ 2c_{GG}\,\frac{m_d-m_u}{m_d+m_u} + k_u-k_d
    + \frac{m_a^2}{m_\pi^2-m_a^2}\,\Delta c_{ud} \right]\! ,
\end{aligned}
\end{equation}
where $k_q$ are the ALP couplings to right-handed quark currents in (\ref{Leff}). We omit a contribution with $(\rlap{\hspace{0.2mm}/}{p_\pi}-\rlap{\hspace{0.2mm}/}{p_a})$ inside the spinor product, which is proportional to the electron mass. For the default choice of the $\kappa_q$ parameters, the term involving $\Delta c_{ud}\equiv c_{uu}-c_{dd}+2c_{GG}\spac\frac{m_d-m_u}{m_d+m_u}$ in the second line is due to ALP--pion mixing. For the QCD axion or a light ALP with $m_a^2\ll m_\pi^2$ this contribution is negligible. In ``pion-phobic axion models'' \cite{Krauss:1986bq} one tunes the couplings $c_{GG}$, $k_u$ and $k_d$ in such a way that the amplitude (\ref{eq:21}) vanishes. This tuning is unnatural, because the couplings $k_q$ change under scale evolution whereas $c_{GG}$ is scale invariant \cite{Bauer:2020jbp}.

Our model-independent predictions in (\ref{resu}) and (\ref{eq:21}) can be compared with results obtained in the context of specific axion models. In the ``variant-axion models'' the coupling parameters in the effective Lagrangian (\ref{Leff}) are obtained as $c_{GG}=-\frac{N}{2}(x+\frac{1}{x})$, $k_u=z$, $k_d=k_s=\frac{1}{x}$ and $k_U=k_D=k_S=0$, where $z$ and $x$ are the Peccei--Quinn charges of the right-handed up and down quarks, and $N$ is the number of up-type quarks with the same charge as $u_R$. With these identifications, our result (\ref{eq:21}) agrees with eq.\spac\spac(4.1) in \cite{Bardeen:1986yb}, and our result (\ref{resu}) agrees with eq.\spac\spac(4.66) upon setting $m_a^2=0$, apart from some subleading corrections of ${\cal O}(m_\pi^2/m_K^2)$. For the ``short-lived axion model'' the relevant couplings are $k_u=-2c_{GG}=1$ and $k_d=0$, and with these values our result (\ref{eq:21}) agrees with a corresponding relation obtained in \cite{Krauss:1986bq}.

In summary, we have present a consistent implementation of weak decay processes involving an axion or axion-like particle in the context of the chiral Lagrangian. We have pointed out that previous calculations have neglected to include important weak-interaction vertices involving derivative couplings of the ALP, which as shown in (\ref{master}) arise when the relevant chiral quark currents are derived from the Noether procedure. Other phenomenological treatments based on the notion of ALP--meson mixing have omitted several relevant contributions. In particular, we find that that $K^-\to\pi^- a$ branching ratio is about a factor 37 larger than the prediction obtained in \cite{Georgi:1986df}, which has important phenomenological consequences. We have derived the model-independent expressions for the $K^-\to\pi^- a$ and $\pi^-\to e^-\bar\nu_e\spac a$ decay amplitudes, including all relevant ALP couplings and the effects of the ALP mass. The methods we have developed can be applied to a variety of other low-energy observables of phenomenological interest.

{\em Acknowledgements:\/} 
M.N.\ thanks Gino Isidori, the particle theory group at Zurich University and the Pauli Center for hospitality during a sabbatical stay. This research has been supported by the Cluster of Excellence PRISMA$^+$\! funded by the German Research Foundation (DFG) within the German Excellence Strategy (Project ID 39083149).

\end{document}